# Neighbor Joining Plus - algorithm for phylogenetic tree reconstruction with proper nodes assignment.

**Piotr Płoński[1][ξ], Jan P. Radomski[2][ξ]**

[1]Institute of Radioelectronics, Warsaw University of Technology, Nowowiejska 15/19, 00-665 Warsaw, Poland

[2]Interdisciplinary Center for Mathematical and Computational Modeling, Warsaw University, Pawińskiego 5A, Bldg. D, 02-106 Warsaw, Poland

## Abstract

Most of major algorithms for phylogenetic tree reconstruction assume that sequences in the analyzed set either do not have any offspring, or that parent sequences can maximally mutate into just two descendants. The graph resulting from such assumptions forms therefore a binary tree, with all the nodes labeled as leaves. However, these constraints are unduly restrictive as there are numerous data sets with multiple offspring of the same ancestors. Here we propose a solution to analyze and visualize such sets in a more intuitive manner. The method reconstructs phylogenetic tree by assigning the sequences with offspring as internal nodes, and the sequences without offspring as leaf nodes. In the resulting tree there is no constraint for the number of adjacent nodes, which means that the solution tree needs not to be a binary graph only. The subsequent derivation of evolutionary pathways, and pair-wise mutations, are then an algorithmically straightforward, with edge's length corresponding directly to the number of mutations. Other tree reconstruction algorithms can be extended in the proposed manner, to also give unbiased topologies.

## Introduction

Rapidly developing computer techniques constantly generate new sets of challenges for phylogenetic analysis – the ever increasing amount of stored information requests new and better tools for revealing sequences' relationships. A phylogenetic tree is a graph, which has no cycles [1], with nodes representing sequences. The nodes which have more than one adjacent node constitute internal nodal points – corresponding to sequences of numerous offspring; and the nodes with only one adjacent node being terminal leaf nodes – representing sequence without offspring. The differences between sequences are presented as sums of edges' weight between each two corresponding nodes. By assigning a root node, and then by traversing from leaf nodes, through internal nodes, to a root node, the evolutionary pathways can be inferred in a straightforward manner [2].

The phylogenetic tree reconstruction is a NP-hard problem [3]. Therefore, for tree reconstruction various heuristic techniques were proposed – generally they can be divided into four groups of methods. The Maximum Parsimony (MP) [4], which builds multitude of tree candidates, each with assumed minimum sum of distances between nodes. The MP is a non-parametric method, however, it is sensitive to different rates of mutation speed on edges,

---

[ξ] corresponding authors: pplonski@ire.pw.edu.pl, janr@icm.edu.pl

resulting often in wrong tree topologies. The Maximum Likelihood method (ML) [5] is a parametric approach, which assumes a mutation model between sequences. It also builds a set of candidate trees, and then chooses a tree with assumed maximum probability of occurrence as a final solution. The efficiency of this method depends on a correct mutation model selection. The biggest disadvantage of both MP and MP is their computational cost during optimal tree topology searching. To make the searching process faster various heuristic methods were proposed (for a review see [6]). The Bayesian generalization of ML forms the third group methods [7] – the final tree obtained by these techniques is an integral over all possible tree topologies. The MP, ML and Bayesian solutions are computationally very demanding, and usually are used on smaller data sets. Finally, the clustering methods based on distance matrices form an alternative approach, and can quickly build a tree for very large data sets. The most popular among them are the Neighbor Joining (NJ) [8] and UPGMA [9]. The NJ can be seen as a classic algorithm in the field of tree reconstruction, as it is one of the most frequently used techniques [10]. It was carefully examined by Attenson [11], who suggested the NJ-classic as a performance benchmark for testing new algorithms [12].

Major algorithms of the phylogenetic tree reconstruction described so far, besides having problems in attaining acceptable efficiency in reconstructing correct tree topology, share the same limitations for presenting reconstructed tree topologies. The resulting trees have only their leaves labeled, which can be interpreted as assuming *a priori* that in every analyzed set there are only sequences without offspring. Therefore, the resulting graph is always a binary tree. Which in turn hinders the subsequent reconstruction of evolutionary pathways, making it much more tedious than necessary. These limitations are of purely algorithmic nature, and can be circumvented by the method we propose here. In this work we present an extension to the NJ classic method, which can properly handle nodes assignment in a tree, by representing the sequences with offspring as labeled internal nodes, and sequences without offspring as leaves. We have called it Neighbor Joining Plus (NJ+). The approach lifts the limitation that resulting tree is constructed in a binary manner. The NJ+ represents each edge length as the number of mutations between connected node sequences.
To our best knowledge, currently the only other available algorithm of phylogenetic reconstruction not burdened with the afore mentioned limitations is the goeBURST [17, 18]. It would be of interest to compare the results between these two approaches directly, however, at present goeBurst does not provide possibility to either count mutations between nodes, or at least to output the resulting graphs in the Newick format.

In the next section the algorithm details are presented. After that, the performance of the NJ+ algorithm is examined on synthetic, as well as on some real data sets, and then compared to the NJ classic results using the Robinson-Foulds (R-F) distance [13]. The comparison of the NJ+ with the corresponding NJ results show that for data sets with only terminal sequences (without offspring), and binary trees, it results in the same tree topology, but with a better accuracy of edge lengths estimation. On data sets with both sequence types (with, and without offspring) and non-binary true trees, the NJ+ provides much improved tree topology reconstructions, including better edge lengths, than the classical NJ. It has no limitations to handle only binary trees. Recently we have proposed the QPF algorithm [14] for translation of biased trees (trees with only their leaves labeled, and binary topology) to provide correct unbiased topologies. The resulting tree topologies obtained by the NJ+ algorithm significantly facilitate a fully automatic, straightforward analysis of evolutionary pathways, and enable sequence mutation direct counting on a pair-wise, as well as on a cumulative basis.

## Methods and Materials

The distance $D(i,j)$ between two sequences $i$ and $j$ is designated as the number of differences between their nucleotide chains. The distance between two nodes $(u,v)$ in the tree will be denoted as $d(u,v)$. The flow of NJ+ algorithm is similar as in the classic NJ approach, with an addition of few extra steps. At first the number of nodes is equal to the number of sequences, so $d(i,j) = D(i,j)$, the distances between nodes are stored in the matrix $d$. The cost matrix $Q$ is then computed:

$$Q(i,j) = (r-2)d(i,j) - \sum_{k=1}^{r} d(i,k) - \sum_{k=1}^{r} d(j,k), \tag{1}$$

where $r$ is the number of still not merged nodes. Then from the $Q$ matrix the entry with minimum value is found, let's note its indexes as $f,g$:

$$(f,g) = argmin(Q). \tag{2}$$

In the classical NJ algorithm a new node $u$ is created, and distances $d(f,u)$ and $d(g,u)$ are computed. In our algorithm NJ+ the new node will be created only after we would decide about its creation in the subsequent steps, however the distances $d(f)$ and $d(g)$ are computed same as in the NJ-classic:

$$d(f) = \frac{1}{2}\left(d(f,g) + \frac{\sum_{k=1}^{r} d(f,k) - \sum_{k=1}^{r} d(g,k)}{r-2}\right), \tag{3}$$

$$d(g) = d(f,g) - d(f).$$

Then the distances $d(f)$ and $d(g)$ are rounded in the way presented below, to assure that edge weight will always represent a full mutations length:

$$d(f), d(g) = DIST\_ROUND(d(f), d(g)). \tag{5}$$

After computing the distances and determining the tree topology, the node character is inferred – that is whether the node will be treated as a leaf or as an internal node. In the NJ-classic the nodes $f$ and $g$ are always assumed to be leaf nodes. The description how we decide on node's character is presented below. In the case of two leaf nodes the algorithm is the same as in the NJ-classic – the new node $u$ is created, and nodes $f$ and $g$ are joined into the $u$ node. The distance between the nodes $u$ and $f$ and $g$ are designated as:

$$d(u,f) = d(f), \tag{6}$$
$$d(u,g) = d(g).$$

The distance between the new node $u$ and the other (not yet added to the tree) node $k$ is computed as:

$$d(u,k) = \frac{1}{2}(d(f,k) + d(g,k) - d(f,g)). \tag{7}$$

Then the nodes *f* and *g* are removed from the *d* matrix, and the new node *u* is added into the *d* matrix instead. So far the steps taken were identical to the NJ-classic approach. Note, that the newly created node *u* has not been assigned sequence's label. In a case that there is one leaf node and one internal node, we need to extend the classic NJ algorithm behavior by adding some new steps. Let's assume that *f* is an internal node, and *g* is a leaf. Then, the node *g* is joined with the node *f* and it is removed from *d* matrix. For a case of the *f* being a leaf and the g an internal node, the steps will be analogous. The last possible configuration of node types to consider will be when both the *f* and the *g* are internal nodes. For such a case we'd need to decide, during their labeling, how we will place them into the tree. If both of them are unlabeled, or if the *f* is labeled and the *g* is unlabeled, then we join with the *f* all the nodes previously joined with the *g*, and then remove *g* from the distance matrix *d*. Notice that *g* will not be present in the final tree, as it was only used as an auxiliary node, to temporarily hold already joined nodes – as in each step only two nodes are being joined. By introducing this step we have dispersed with the limitation for a number of simultaneously joined nodes. In a situation when the *g* is a labeled node, and the *f* is an unlabeled one, we will proceed analogously. In the case when both *f* and *g* are labeled we take the same steps as for the both nodes being terminal leaves. The algorithm steps are summarized in a pseudocode as follows:

```
Repeat until there will be only one node left in the distance
matrix d.
   1. Compute the cost matrix Q.
   2. Find an entry (f,g) with minimum value in the matrix Q.
   3. Compute distances d(f) and d(g).
   4. Round distances d(f) and d(g) to full-mutation distances.
   5. Determine the nodes' f and g character.
   6. Merge nodes:
         a. If f and g are leaf nodes:
              i. Create a new node u;
             ii. Join nodes f,g with new node u;
            iii. Set d(u,f) = d(f) and d(u,g) = d(g);
             iv. Remove f and g from the distance matrix d;
              v. Add u to the distance matrix d, and update the
                 matrix d values.
         b. If f is an internal node and g is a leaf:
              i. Join node g with node f;
             ii. Remove g from the d matrix.
         c. If g is an internal and f is a leaf:
              i. Join node f with node g;
             ii. Remove f from d matrix.
         d. If f and g are internal nodes:
              i. If f is labeled and g is unlabeled or if both
                 are unlabeled:
                   1. Move to f all previously joined nodes of
                      g;
                   2. Remove g from d matrix.
             ii. If g is labeled and f is unlabeled:
                   1. Move to g all previously joined nodes of
                      f;
                   2. Remove f from d matrix.
            iii. If both are labeled:
                   1. follow the steps same way as for the 6a.
```

**Distance rounding**

The distance between two sequences is an integer representing a number of differences between their nucleotides chains. However, the distances computed in the equation (3) are real numbers. To assure that all edges in the tree represent a number of mutations between sequences we round their distances. The DIST_ROUND function is described in the pseudocode as follows:

```
d(f), d(g) = DIST_ROUND(d(f), d(g)):
Sum = d(f) + d(g);
// {d(f)} is a fractional part of d(f)
// round(d(f)) is the usual round function
if( {d(f)} > {d(g)} ) then
    d(g)=round(d(g))
    d(f)=Sum-round(d(g))
else
    d(f)=round(d(f))
    d(g)=Sum-round(d(f))
end;
```

Let's see how it works on an example, for *d(g)=3.2* and *d(f)=3.4* the DIST_ROUND function will return *d(g)=3* and *d(f)=4*. It is important to keep in mind that it is a non-standard rounding routine.

**Determination of node's character**

The crucial part of NJ+ algorithm is to determine a character of the node. Node can be in one of the three states: (i) leaf, (ii) internal node with label, (iii) internal node without label. This respectively corresponds to: (i) sequence without offspring, (ii) sequence with offspring that is present in the dataset, (iii) missing ancestor sequence[s] with offspring not present in the analyzed dataset. Determination of node's character is decided by evaluating two conditions during internal nodes searching. First, the node's edge length *d(f)* and *d(g)* is examined, if edge length is smaller than one (which corresponds to exactly one mutation), then the node is treated as an internal node, otherwise it is a leaf. If, by fulfilling this condition, nodes are assigned to be leaves, the second condition is used to ensure that any of such node is not an internal node. Second condition checks triangular equality between three node sequences. Let's assume that the *f* and *g* are labeled nodes. When *f* or *g* have not a label, the label for them is taken from theirs joined nodes – as the label of the closest labeled node. When, as in creating a third labeled node, we are about to select the node *h*, which is either the closest to the *f* or to the *g*, and which is still present in the distance matrix *d* – then for the candidate selected nodes we have their corresponding sequences. Lets denote them as F, G, H. We need then to check whether their distances will satisfy the equation:

$$D(F,H)=D(F,G)+D(G,H). \quad (8)$$

If this equation is satisfied then node G is an internal node, and the node F will be joined with G. If equation (8) is not satisfied the analogous condition is checked, which assume the F to be a possible internal node. The assignment of the label for an internal node is made by considering label existence of a node in question.

**The choice of distance metric**

In the proposed method we have chosen as a distance metric the number of mutations between sequences as it guarantees that when triangle equality is satisfied: the distance between mother and granddaughter will be equal to the sum of distances: between mother and daughter, and between daughter and granddaughter (with the assumption that there were no multiple mutations on the same position). The selection of the distance metric between sequences that satisfy triangle equality is necessary to present evolution of the sequences, as we want to distinguish between the two types of phylogenetic trees:
  (i) clustered graph, which shows sequences that are similar near each other in the tree, and places divergent sequences far from each other – from such a tree the evolutionary events (like mutations) can not be read directly;
  (ii) evolutionary tree, which not only shows similar sequences near, and different sequences far in the tree, but also allows to read evolutionary events from the tree.

The clustering graph has wider meaning than evolutionary tree. The assumption of the triangle equality is embedded in NJ itself. There was many attempts to explore the phenomena of assigning various of cost metrics in NJ[19]. However, we here we concentrated on examining more closely the distance merging equation (3). If triangle equality is satisfied between all sequences, then each distance can be rewritten as a sum of partial distances:

$$d(i,j) = d(i,1) + d(1,2) + d(2,3) + ... + d(n,j)$$

where 1,2,3,…n are sequences that form theh between i and j in the **known** tree. If we substitute this into (3) it will result in:

(i) i case if *f* is an offspring of *g* :

$$d(f) = \frac{1}{2}\left( d(f,g) + \frac{\sum_{k=1}^{r-2} d(f,g)}{r-2} \right) = d(f,g),$$

$$d(g) = d(f,g) - d(f) = 0.$$

(ii) i case of *f* and *g* are offspring of another node *u:*

$$d(f) = \frac{1}{2}\left( d(f,g) + \frac{\sum_{k=1}^{r-2} d(f,u) - \sum_{k=1}^{r-2} d(g,u)}{r-2} \right) = \frac{1}{2}\left( d(f,u) + d(u,g) + d(f,u) - d(g,u) \right) = d(f,u),$$

$$d(g) = d(f,g) - d(f) = d(u,g).$$

The above equations ensure that sequences will be presented in evolutionary manner on the tree, and the distances between them will represent an evolutionary distance. In case of other metrics, for example Jukes-Cantor (JC) or Kimura two-parameter model (K-80), the tree will be presented in a clustered manner. The NJ+ will still work with both the JC or K-80 metrics, however, it will act then the same as NJ-classic algorithm, yielding the NJ-classic tree topology (a clustered tree) from which evolution can not be read. Therefore, here we take into consideration only mutations to represent evolutionary steps. Possible recombination events are not considered. For tree visualization the Dendroscope package was used [16].

**Results and discussion**

The proposed NJ+ method was tested on several artificial data sets, as well as on the two real data sets. These tests will be presented in order of an increasing difficulty for tree reconstruction. The NJ+ method is compared here only with the NJ-classic method because: first, it is extension of this method, and second, we are adhering strictly to the suggested benchmarking scheme of Attenson [11]. Of course, should the approach presented here be extended also to other methods for phylogenetic tree reconstruction, then a comparison with other methods would be interesting as well.

First a demonstration of differences on a very simple artificial set is presented. Short nucleotide chains were used (**Fig. 1A**), mutated in a stepwise manner from the starting "AAAAAAAA" string. The **Fig. 1B** shows a typical NJ-classic tree, and **Fig. 1C** a corresponding tree resulting from the NJ+ method. Internal nodes in the NJ-classic trees play only an auxiliary, a strictly technical role, and the resulting tree is of binary character – obscuring ancestor-descendant relationships between sequences. In contrast, the tree reconstructed by the NJ+ assign the internal, ancestral nodes directly, assigning as leaves only sequences with no offspring. The next test was dealing with more difficult to reconstruct synthetic data sets – obtained by joining several small sub-trees. The basic small tree is shown in the **Fig. 2A**; comprising of eleven nodes, from which five are internal nodes, and the remaining six are leaves. Notice, that it is not a binary tree, as the numbers of adjacent nodes varies. Building of the larger tree was done in a stepwise manner, by replacing some leaf node[s] with yet another basic small tree. The tree growing process was started from the lowest leaf node number – an example of a tree with 21 nodes is presented on the **Fig. 2B**. The overall node numbers in the generated trees were S = {11, 21, 51, 71, 101}. The distances between adjacent nodes in the tree were all constrained to one, corresponding to a single mutation between sequences, and random mutations were introduced, starting from the sequence labeled 's.1'. As a starting point a sequence with only 'A' nucleotides was taken each time. We generated sequences with varying chain lengths L = {10, 50, 100, 200, 500}. As a criterion of the reconstruction accuracy the Robinson-Foulds distance [13] was used – it estimates the percent of mismatching edges between two trees (R-F). To make comparison more robust we measured the absolute differences between matching edges, and took the whole edge's length in cases of miss-matching ($E_{diff}$). Comparisons were done between the synthetic and reconstructed trees. The results comparing the NJ-classic and NJ+ methods are presented in the **Figure 3**. It can be observed that with increasing chain lengths the number of incorrectly reconstructed edges decreased exponentially. However, for the NJ-classic method this improvement has reached a plateau at 33.3%, whereas the NJ+ method reconstructed almost all edges correctly. This is a direct consequence of the NJ-classic inability to cope with the number of neighbors for internal nodes higher than three – resulting in a significant bias of the NJ-classic reconstruction accuracies.

Analyzes of the real sequence data sets need to deal with situations where significant ratios of data gaps are present, as usually most sequences from evolutionary processes are not isolated. The question of how well a given phylogenetic reconstruction algorithm is able to cope with such missing ancestors (MAs) touches an area of quite specific challenges. Therefore, the next suite of tests was performed on the data sets already described, but with addition of the random removal of some fraction of all sequences. They were removed only from internal nodes, and the percent of sequences remaining from the original data was varied: R = {55%, 64%, 73%, 82%, 91%}. The resulting R-F differences between the synthetic tress, with an increasing amount of gaps, *vs.* the NJ-classic, and the NJ+ trees are shown in **Figure 4**. The comparisons were measured on data sets with sequence chain length equal to 200 nucleotides. It can be observed that the efficiency of tree reconstruction increase for NJ-classic with the number and the size of gaps. This is because NJ-classic can not cope with internal nodes, and with increased removal of more internal nodes from a true tree, the better reconstruction can be gained. On the other hand, the efficiency of tree reconstruction for NJ+ is slightly decreasing with gap size growing, which is an expected behavior as information from data set is getting more and more scarce. For sets corresponding to the synthetic trees containing only leaves (R=55%), their reconstruction efficiencies by the NJ-classic and the NJ+ were similar. It was not the same because in these cases we reconstructed the leaf-only non-binary trees.

Sanson *et al.* [15] described a very interesting laboratory-controlled evolution experiment in which all the evolving replicates of *Tryponosoma cruzi* were known, isolated and sequenced. What is more, also the evolutionary relationships between the all sequences were known, as the true (laboratory) tree was known too. The whole data set contained 31 sequences with the chain lengths of 2234 nucleotides [accession numbers: GenBank AF288660, GenBank AF359467- AF359496]. The derived evolutionary tree of *T. cruzi* is presented in **Figure 5**.

First we compared performances of the NJ-classic and the NJ+ on synthetic data sets created to mimic the evolution in the *T. cruzi* laboratory tree, generating data sets with different chain lengths: L = {223, 2234, 6702}. As a starting point we took the original sequence '1.1'. Additionally, the performance was tested on subsets with different number of sequences removed from the original 31 sequences: R = {0, 3, 6, 9, 12, 15}. The removal was done only on internal nodes, so for the highest R we had tree with all sequences present only as the leaf nodes. The performance comparison of NJ-classic and NJ+ with laboratory tree is shown in the **Figure 6**. The edge length accuracy reconstruction was much higher for the NJ+ than for the NJ-classic. The NJ+ resulted in the smaller $E_{diff}$ errors for all sequence lengths, and for various numbers of removed sequences. The error in proper edge length reconstruction increased with the increasing number of removed sequences. Diminishing a number of sequences affects the information encoded in a set – crucial for a proper tree reconstruction. The number of correctly reconstructed edges for the NJ+ and NJ-classic increases (the R-F value decreases) when the number of removed sequences increases. This is very surprising and non-intuitive, as with removing information from a data set the efficiency in tree reconstruction will decrease on $E_{diff}$ plots. This is observed for NJ-classic because with sequence removing we are closer to binary, leaf labled only tree, for which NJ-classic was designed. Therefore it is gaining a better reconstruction efficiency when challenged with sequences removal at all chain lengths. However, for NJ+ the same behavior was present only for the chain's lengths 223 and 2234, but not for the chains of 6702 nucleotides. For NJ+ we didn't observe same effect as for NJ-classic – the higher probability to observe several mutations on a given position when the chain length decreases (and/or when the number of sequences in the set increases). It is worth noting, that independently of chain length, in cases when we consider trees in which the only labeled nodes are the leaves, the NJ+ and the NJ-classic reconstruct correctly the same number of edges, as indicated by the same R-F values (this is because reconstructed true trees were the leaf-labeled only, binary trees). However, the NJ+ still yields a higher accuracy of edges' length reconstruction.

The same experiment was then performed on *T. cruzi* laboratory data set. The results are presented in the **Figure 7**. The shapes of graph curves are strikingly similar to the corresponding curves shown on **Figs. 6A** and **6B**. As this suite of tests was done for the chain lengths of 223 rather than 2234 nucleotides, we assume that such similarity was caused by a high number of multiple mutations present on the same chain position, which would make the laboratory data set similar to sets of synthetic sequences created, and yet on chains ten times shorter (for details see [15]). We obtained 93.3% rate of correctly reconstructed mutations by using all sequences from *T. cruzi* data set. The **Figure 8** shows the trees reconstructed by the NJ-classic, and the NJ+ for the laboratory data set. The tree generated by the NJ-classic contains 61 nodes, and in the NJ+ one – 34 nodes (in the analyzed set there were 31 sequences). The tree obtained by NJ-classic is presented as cladogram in the **Fig 8A**, where internal nodes without labels can be clearly observed. There was no such unambiguous observation in case of the NJ-classic phylogram (**Fig. 8B**), where the leaf nodes (which represent sequences with offspring) are very close to internal nodes. However, they are situated on rather short edges. In the NJ+ cladogram (**Fig. 9D**) it can clearly seen that all internal nodes are labeled. This is also visible in NJ+ phylogram (**Fig. 8D**). Here the both threes reconstructed by the NJ-classic and the by NJ+ were binary, as the true laboratory tree to be reconstructed was binary also.

In order to illustrate the problem of inadequacy of existing phylogenetic trees to cope with ancestor sequences producing multiple (that is not binary) offspring, we have culled a small fragment of just few branches (67 sequences total) from a large tree for the 2009 pandemic H1N1 virus hemagglutinin. The input tree, built from 3243 unique sequences of 1701 nucleotides each, contained many places where it was clear that respective mother sequences had multiple offspring (in the most striking case 41 unique daughter sequences). Each offspring sequence was distinct, and differed by just a SNP from its ancestor – therefore there were no ambiguity as to their respective ancestry (see Table 1). The trees, reconstructed for this small set of 67 sequences, by the NJ-classic, and the NJ+ are presented on the **Figure 9**. The **Table 1** contains the distance matrix for the some example sequences highlighted for comparison purposes (the same color is used for highlight corresponding sequences in the trees of the **Fig. 9**). In the NJ-classic tree there were 133 nodes, and in the NJ+ tree – 67 nodes (every NJ+ node was labeled). The drawback of binary constraint of the NJ-classic is well observed in NJ-classic cladogram (**Fig. 9A**). It can't be as clearly observed in the case of NJ-classic phylogram (**Fig. 9B**), as distances between auxiliary internal nodes are all zero. Because of binary limitations in the NJ-classic tree there are several edges distancing each mother and daughter pair nodes. In the NJ+ there is always a spacer of exactly one edge between each daughter and mother pairs of sequences.

## Conclusions

The presented NJ+ method yield trees with proper nodes assignment. There is no constraint that the resulting trees will be binary only. The length of edges represents the number of mutations between sequences. The performed tests on synthetic and real data sets confirm that the NJ+ method reconstruct trees with higher accuracy than the classic NJ algorithm. Obtaining evolutionary pathways and pair-wise mutations from the reconstructed trees is then straightforward, and facilitates further phylogenetic analyses. Other distance based, tree reconstruction algorithms can be extended in the presented manner as well, to give unbiased, unambiguously labeled topologies.

The NJ+ algorithm was implemented in C++ as a command-line application. It can be obtained from authors by an e-mail request.

## Acknowledgements

We would like to thank Pat Churchland for looking over the English. This work was partially supported by the EU project SSPE-CT-2006-44405, and also partially supported from the BST/115/30/E-373/S/2012 funds.

**Figures**

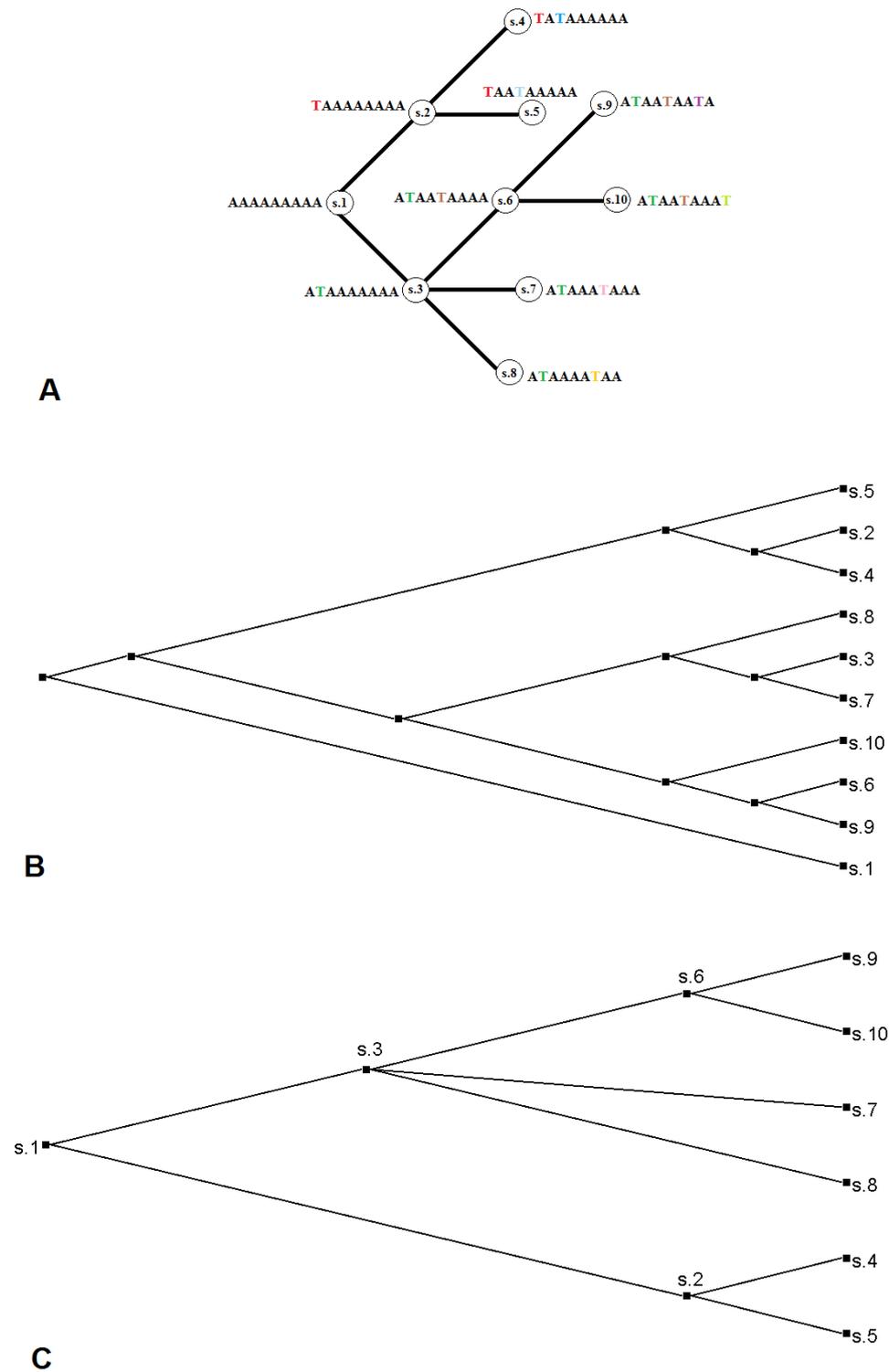

**Figure 1** – Tree reconstruction on simple data set.
Comparison of tree reconstruction between the NJ-classic and NJ+ on a simple artificial data set: (a) true tree with the sequences chains shown near corresponding nodes, (b) tree reconstructed with the NJ-classic algorithm, (c) tree reconstructed with the NJ+ method.

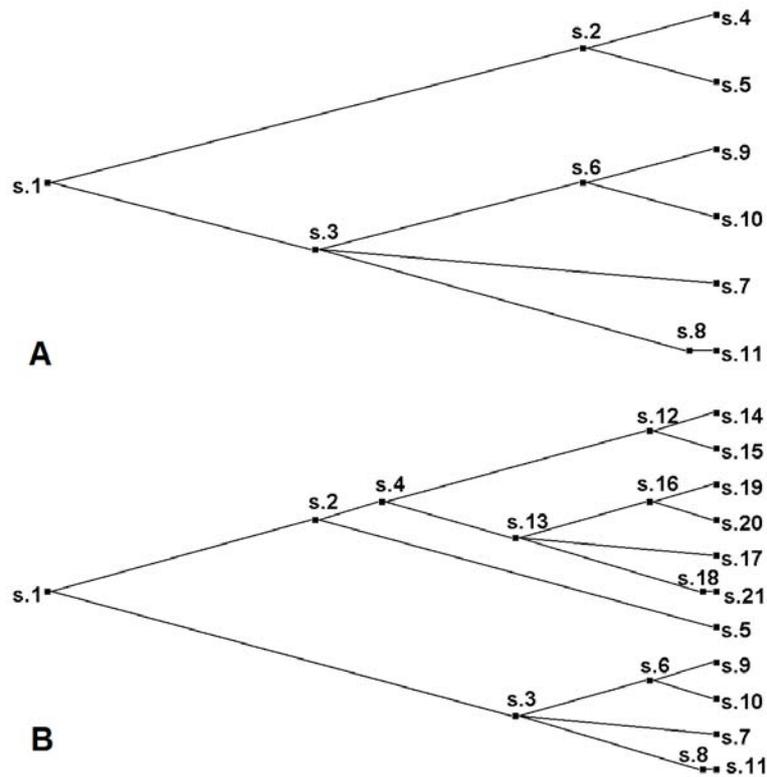

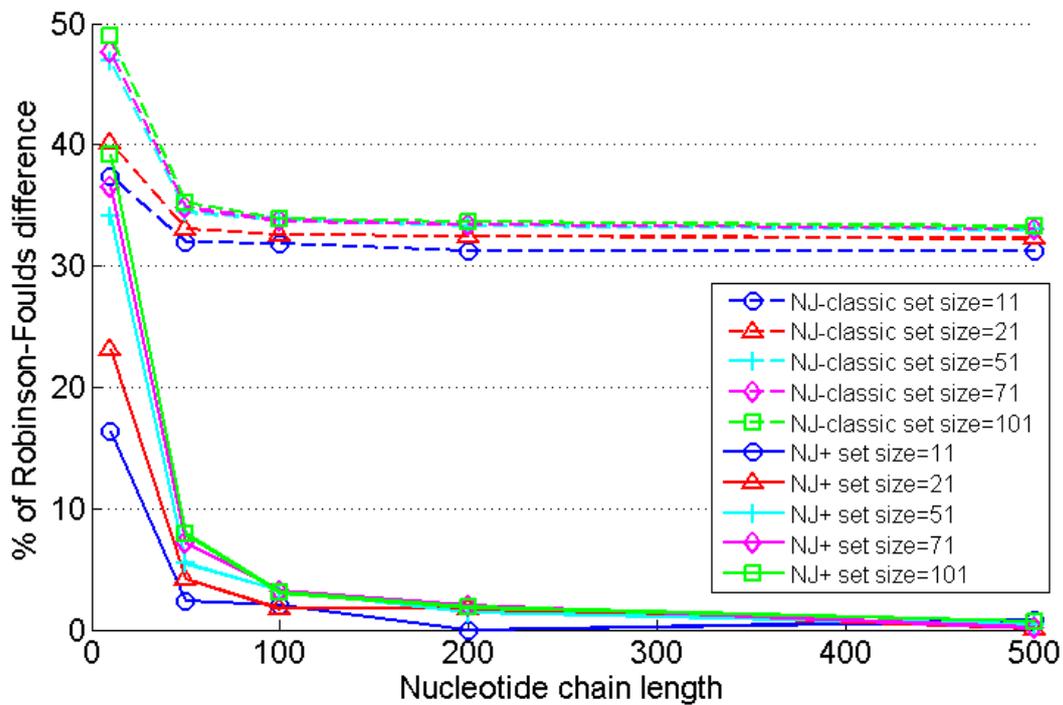

**Figure 2** – Base trees used for artificial data set generation.
(a) the basic tree from which bigger trees were constructed, (b) an example tree with 21 nodes.

**Figure 3** – Tree reconstruction efficiency in dependence on nucleotide chain length.
The comparisons between synthetic trees *vs.* their reconstruction accuracies for the NJ-classic, and the NJ+ methods – plotted as mean values of 50 repetitions.

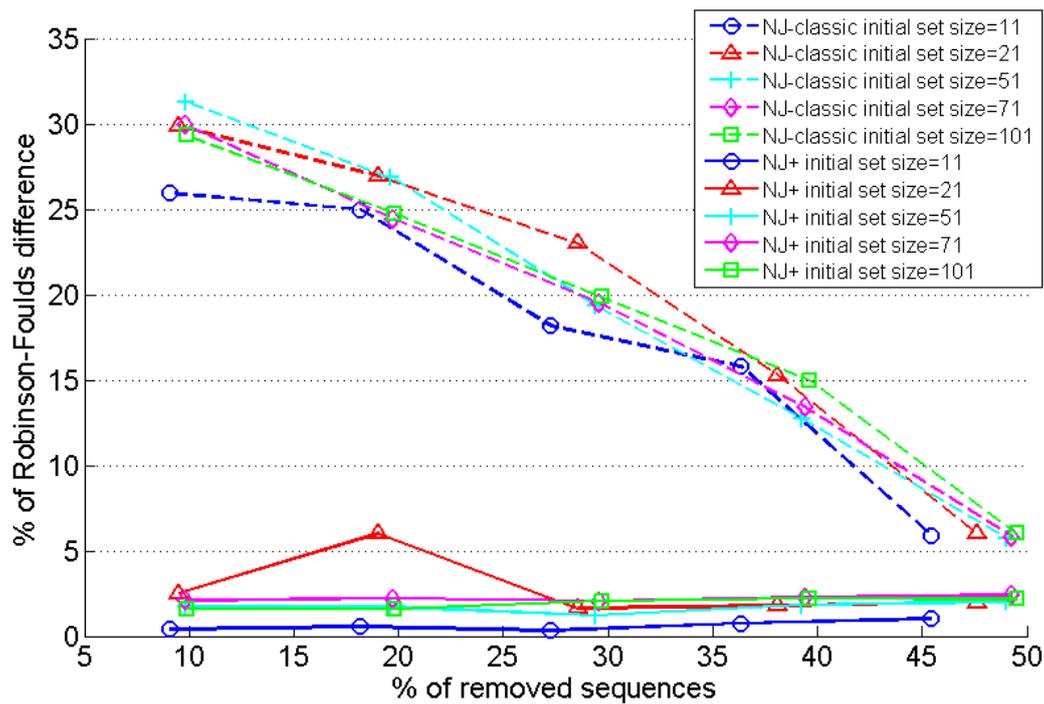

**Figure 4** - Tree reconstruction efficiency in dependence on percent of removed sequences.
The comparison of tree reconstruction efficiency, shown as the R-F differences between synthetic trees *vs.* their reconstruction accuracies: for the NJ-classic, and the NJ+ methods. The comparisons were measured for different number of gaps in the data set, the sequence chain length was 200 nucleotides. Results are mean values of 50 repetitions.

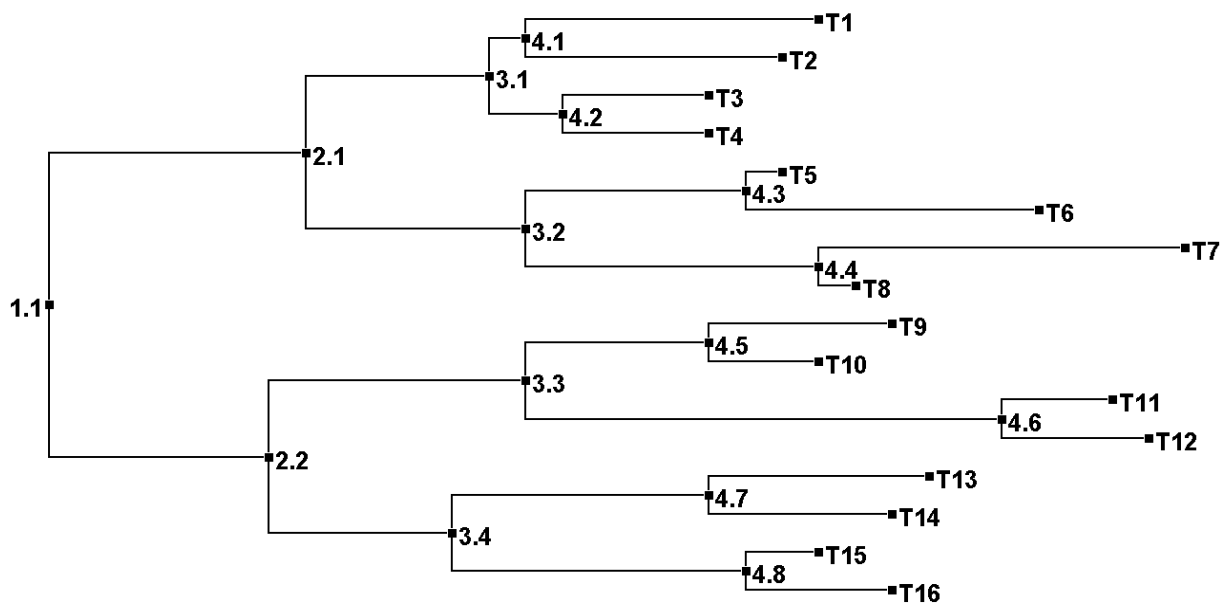

**Figure 5** - The true tree for the *T. cruzi* data set.
The tree for the *T. cruzi* laboratory data set as given by Sanson *et al.* [15].

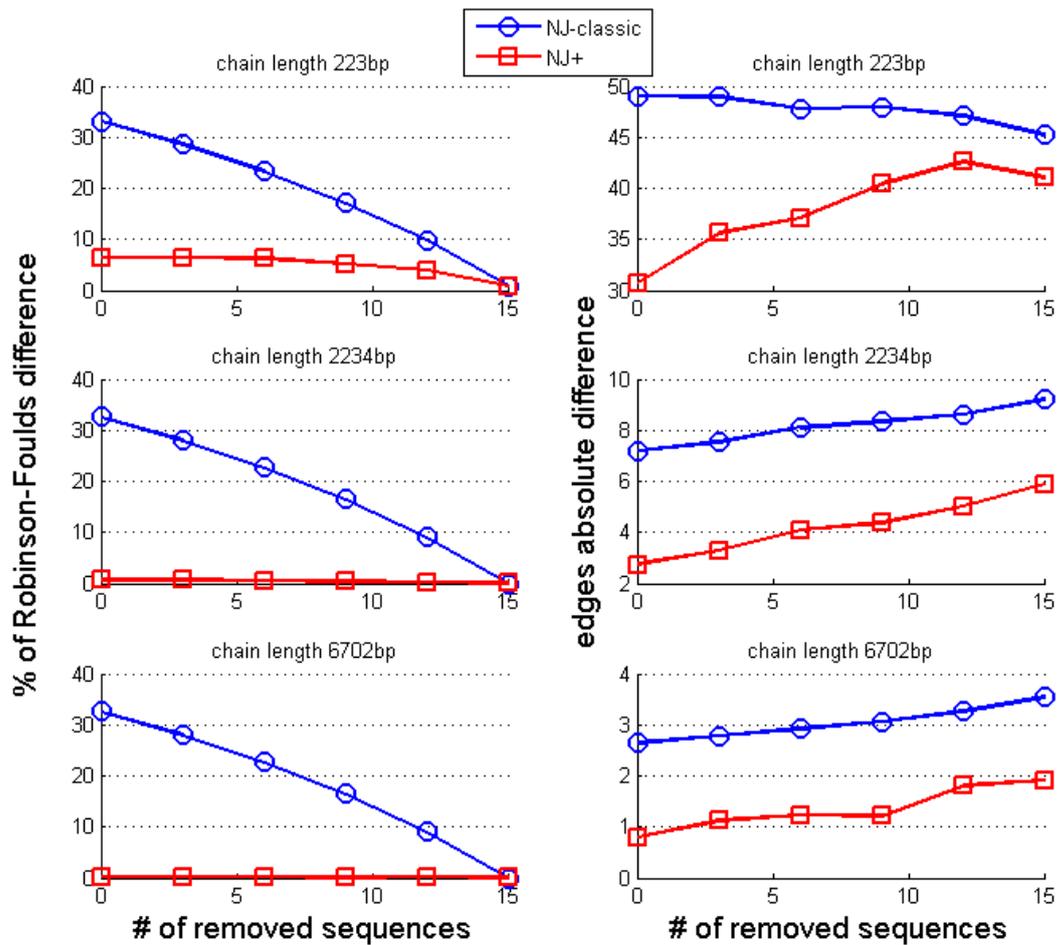

**Figure 6** - Tree reconstruction efficiency on artificial sets that mimic evolution in *T. cruzi* set. The comparison of the NJ-classic *vs.* the NJ+ between synthetic date sets that mimic evolution of the *T. cruzi* true tree, generated with different chain lengths. The results are averaged over 50 repetitions at each data point (each time new starting data set was generated).

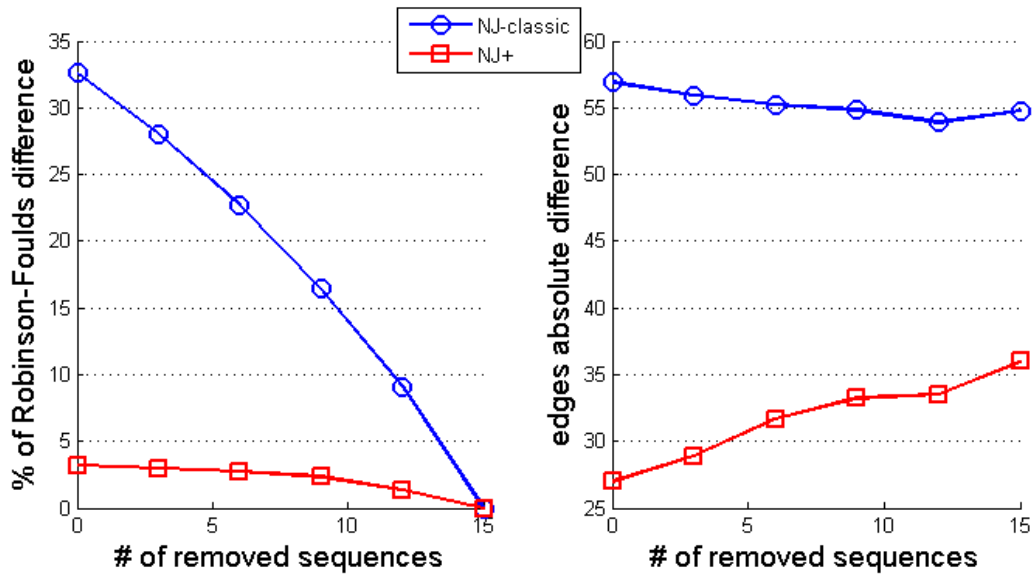

**Figure 7** - Tree reconstruction efficiency on real *T. cruzi* set.
The comparison of results from the NJ-classic and the NJ+ for the laboratory data set of *T. cruzi* sequences – the tests were performed on different numbers of removed sequences For R={3, 6, 9, 12} there were 50 repetition of the experiment with different gap's drawings, for R={0, 15} there was one repetition as there is only one combination of sequences.

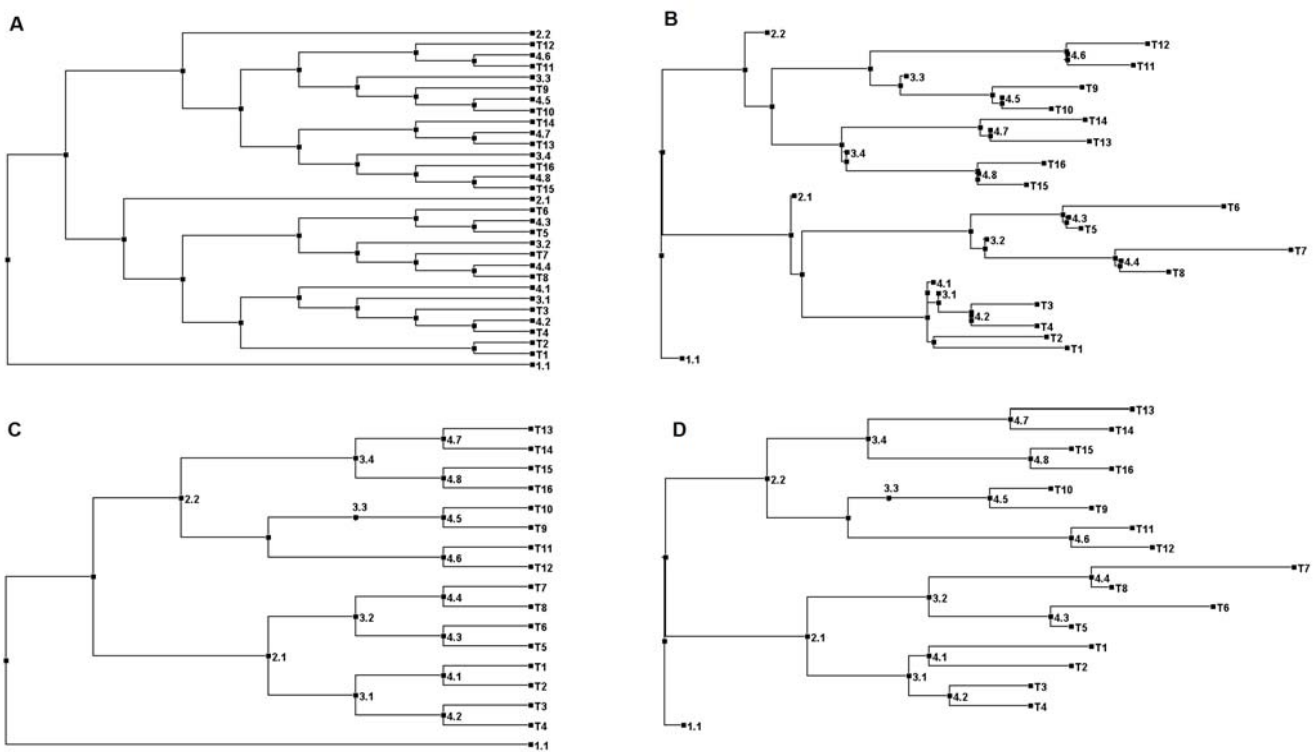

**Figure 8** – *T.cruzi* trees reconstructed with NJ-classic and NJ+.
The *T. cruzi* tree as reconstructed by the NJ-classic (panels A and C), and the NJ+ methods (panels B and D) – all the sequences of the laboratory data set were used.

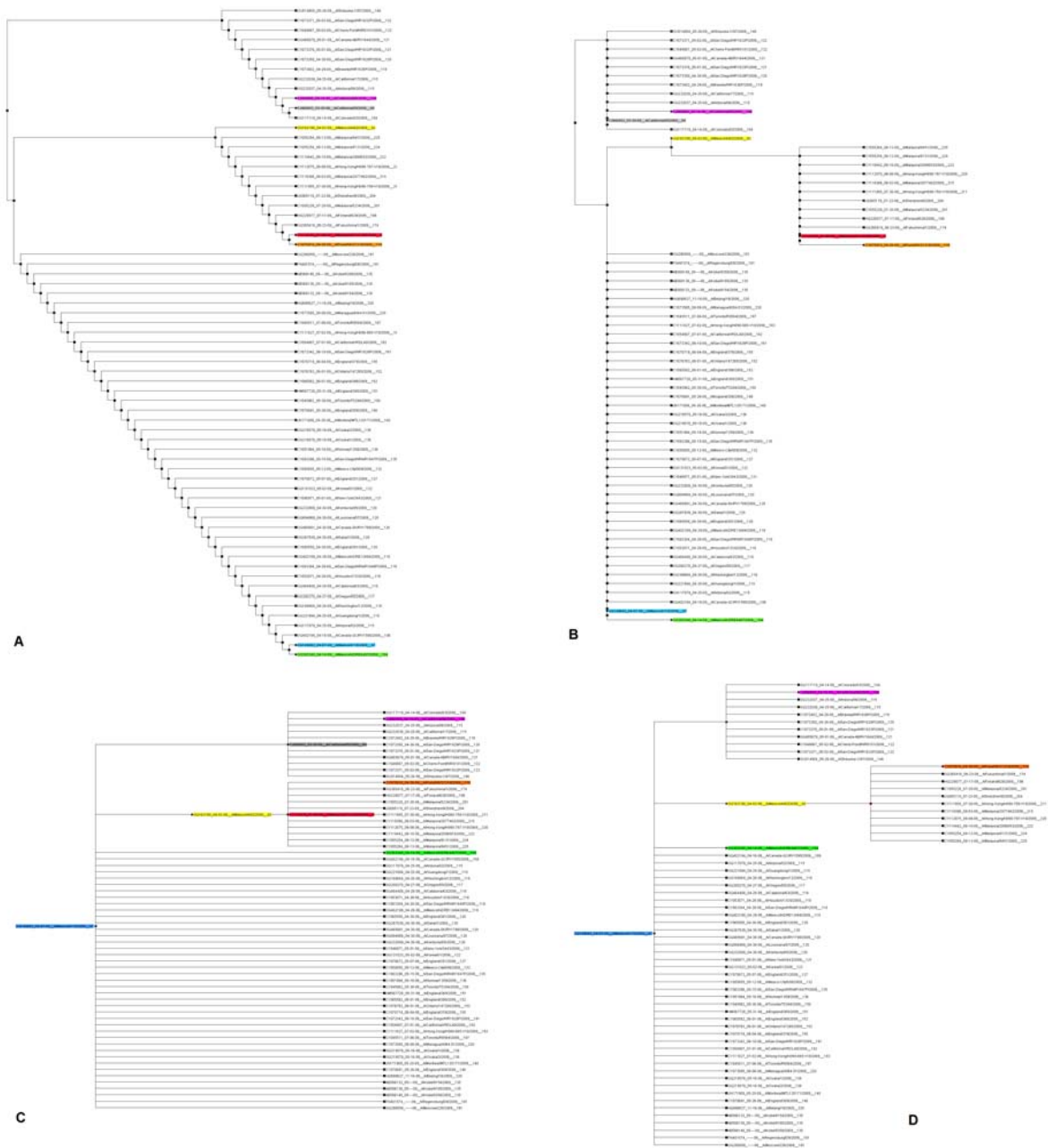

**Figure 9** – Multiple offspring example reconstructed with NJ-classic and NJ+.
The reconstructed trees for a small subset of 2009 pandemic H1N1 virus hemagglutinin influenza of 67 sequences: (panel A) NJ-classic cladogram, (B) NJ-classic phylogram, (C) NJ+ cladogram, (D) NJ+ phylogram.

# Tables

| accession # | GQ149692 | GQ303340 | GQ162190 | CY119330 | CY075910 | FJ966952 | FJ966960 |
|---|---|---|---|---|---|---|---|
| GQ149692 | 0 | 1 | 1 | 3 | 4 | 1 | 2 |
| GQ303340 | 1 | 0 | 2 | 4 | 5 | 2 | 3 |
| GQ162190 | 1 | 2 | 0 | 2 | 3 | 2 | 3 |
| CY119330 | 3 | 4 | 2 | 0 | 1 | 4 | 5 |
| CY075910 | 4 | 5 | 3 | 1 | 0 | 5 | 6 |
| FJ966952 | 1 | 2 | 2 | 4 | 5 | 0 | 1 |
| FJ966960 | 2 | 3 | 3 | 5 | 6 | 1 | 0 |

**Table 1** – Distance matrix between selected sequences from **Fig. 9**.
The matrix of distances between example sequences (*cf.* Fig. 9 for the corresponding trees) – each table cell contains the number of mutations between the respective sequence pairs.